# Barrier Lowering and Backscattering Extraction in Short-Channel MOSFETs

Gino Giusi[(a)], Giuseppe Iannaccone[(b)], Debrabata Maji[(c)], Felice Crupi[(a)]

[(a)]DEIS, University of Calabria, Via P. Bucci 41C, I-87036 Arcavacata di Rende (CS), Italy

(ggiusi@deis.unical.it; crupi@unical.it )

[(b)]DIIEIT, University of Pisa, Via Caruso 16, I-56126 Pisa, Italy ( g.iannaccone@iet.unipi.it )

[(c)] Indian Institute of Technology, Bombay 400076, India (dmaji14@gmail.com)

## Abstract

In this work we propose a fully experimental method to extract the barrier lowering in short-channel saturated MOSFETs using the Lundstrom backscattering transport model in a one sub-band approximation and carrier degeneracy. The knowledge of the barrier lowering at the operative bias point in the inversion regime is of fundamental importance in device scaling. At the same time we obtain also an estimate of the backscattering ratio and of the saturation inversion charge. Respect to previously reported works on extraction of transport parameters based on the Lundstrom model, our extraction method is fully consistent with it, whereas other methods make a number of approximations in the calculation of the saturation inversion charge which are inconsistent with the model. The proposed experimental extraction method has been validated and applied to results from device simulation and measurements on short-channel poly-Si/SiON gate nMOSFETs with gate length down to 70 nm. Moreover we propose an extension of the backscattering model to the case of 2D geometries (e.g. bulk MOSFETs). We found that, in this case, the backscattering is governed by the carrier transport in a few nanometers close to the silicon/oxide interface and that the value of the backscattering ratio obtained with a 1D approach can be significantly different from the real 2D value.

*Keywords*: Barrier Lowering, Backscattering, Device Simulation, C-V Measurements.

# I. Introduction

Due to the continuous downscaling of MOSFETs geometry, improved physical models are needed to accurately study the charge transport in the channel [1-8]. One of the simplest and most successful models was proposed by M. Lundstrom [2] on the basis of the Natori theory for ballistic MOSFETs [1]. In his *backscattering* model, charge transport in the channel is regulated by the injection of the near equilibrium thermal carriers at the top of the source-channel potential barrier (the virtual source). Only a fraction of the injected carriers reach the drain side due to scattering in the channel. The ratio of the backscattered current to the total current injected by the virtual source is the backscattering coefficient.

The backscattering coefficient plays a pivotal role in understanding the scalability of a given technology (material and/or architecture). Since a wide range of technology options are currently under study to meet performance targets as scaling continues, the backscattering coefficient is gaining more and more popularity and experimental procedures for its accurate estimation are strongly required [9-12].

Most of measurements of the backscattering ratio have been done by using the method proposed by Chen [9], where the backscattering is extracted by measuring the saturation drain current at different temperatures [13-17]. However, as it has been discussed by Zilli [18], the method accounts for a number of assumptions which strongly affect the value of the extracted backscattering. A more reliable method has been proposed by Lochtefeld [10, 11], where the saturation inversion charge is obtained from the measurement of the gate-to-channel capacitance corrected for the drain-induced barrier lowering (DIBL). However, because no experimental method exists to extract the DIBL at the specified bias point in the inversion regime, the charge is calculated by using a DIBL extracted in the sub-threshold regime where it is easily calculated as a simple shift of the gate voltage for a constant drain current. Because the DIBL is generally a function of the bias point, the extracted value of the backscattering can be sensibly affected.

Differently, in this paper, we propose a fully experimental method which uses the correct DIBL and it allows the extraction of the barrier lowering directly in the inversion regime, which is of fundamental importance for device scaling, obtaining at the same time an estimation of the backscattering ratio and of the saturation inversion charge. Secondary transport parameters like injection velocity and mean free path can be evaluated as a direct consequence.

The remainder of the paper is divided as stated in the following. In Section II the Lundstrom backscattering model is discussed. In this Section we also extend the model to the case of a two-dimensional (2D) geometry. In Section III the method proposed by Lochtefeld for the backscattering ratio extraction is discussed. In Section IV the proposed method is presented and in Section V it is exemplified through application to 2D quantum corrected device simulations and to measurements on short-channel poly-Si/SiON gate nMOSFETs.

## II. The Backscattering Model

The saturation backscattering coefficient ($r_{sat}$) is defined as the ratio of the negative directed current ($I^-$) at the virtual source ($x_0$), to the positive directed current ($I^+$) at the virtual source (Fig. 1)

$$r_{sat} = \frac{I^-}{I^+} \tag{1}$$

$$I_{D,sat} = I^+ - I^- \tag{2}$$

where $I_{D,sat}$ is the drain current in saturation. The Lundstrom backscattering model in saturation is governed by the equations [3]:

$$I_{D,sat} = qWN_{2D}v_{th}\Im_{1/2}(\eta_{sat})\frac{1-r_{sat}}{2} \tag{3}$$

$$Q_{sat} = qN_{2D}\frac{1+r_{sat}}{2}\Im_0(\eta_{sat}) \tag{4}$$

where $q$ is the electron charge, $W$ is the device width, $N_{2D} = kT\frac{gm_{DOS}}{\pi\hbar^2}$ is the two-dimensional effective density of states with $k$ the Boltzmann constant, $T$ the absolute temperature, $g$ the sub-band degeneracy, $\hbar$ the reduced Planck constant, $m_{DOS}$ the effective electron mass for the density of states of the considered sub-band, $v_{th} = \sqrt{2kTm_C/(\pi m_{DOS}^2)}$ is the average one-dimensional (1D) thermal velocity with $m_C$ the conduction effective electron mass for the considered sub-band, $Q_{sat}$ is the inversion charge per unit of area at the virtual source, $\Im_{1/2}$ ($\Im_0$) is the Fermi-Dirac integral of order one half (zero) and $\eta_{sat} = (E_{FS} - E_1)/kT$ is the energy distance, in units of $kT$, of the populated sub-band ($E_1$) with respect to the source quasi Fermi level ($E_{FS}$). The model is intrinsically one-dimensional (along the channel direction) and for this reason the theory has been developed mainly for thin double gate (DG) devices. Moreover, the Lundstrom model assumes that only one sub-band ($E_1$) is populated. This approximation is good especially for high transverse fields. An empirical approach to take into account multi-band occupation has been proposed by Barral [12] in the case of DG/SOI devices. In this work, to validate the proposed method also in the case of bulk devices, we use a 2D approach for calculating the backscattering ratio.

## IIA. Extension to a 2D geometry

As stated, the backscattering model is 1D and based on a sub-band description, while we need to calculate $r_{sat}$ in a 2D geometry and a quantum corrected semi-classical simulation. To serve this purpose, we propose a straightforward extension of the Lundstrom backscattering model for a 2D semiclassical device. Indeed, in a 2D semiclassical device the position of the virtual source along the longitudinal direction ($x$) depends on the position of the plane along the vertical ($y$) axis. We divide the channel, along the $y$ direction, into thin slices of length $L$ (the channel length) and thickness $\Delta y$ (see Fig. 1). For each slice we find the position of the virtual source $x_0(y)$ and we calculate the local charge density and the local current as

$$Q(y) = q \int_{y}^{y+\Delta y} n(x_0, y')dy' \tag{5}$$

$$I_D(y) = \int_{y}^{y+\Delta y} J_D(x_0, y')dy' \tag{6}$$

where $n$ is the electron concentration and $J_D$ is the current density. Now the 1D model (Eq. 3-4) can be used within the slice, getting the local values for $Q(y)$, $I_D(y)$ and $r(y)$. By solving (1) and (2) within the slice we can calculate also the local values for $I^+(y)$ and $I^-(y)$. Now the 2D backscattering ratio can be calculated as

$$r_{2D} = \frac{\sum_{y'} I^-(y')}{\sum_{y'} I^+(y')} \tag{7}$$

Figure 2 shows the results of device simulation performed on short-channel silicon devices ($L$=70 nm, $t_{ox}$=1.2 nm) biased in saturation. As expected, the currents $I^+$, $I^-$ have a maximum close to the interface (~1nm) where also the charge has the maximum, and decrease rapidly going far from it. $r_{2D}$ is the cumulative 2D backscattering coefficient evaluated through Eq. 7 as a function on the integration depth along the vertical direction $y$. As expected, it can be observed from Fig. 2 that $r_{2D}$ is significantly affected by the charge transport in a region of few nanometers close to the interface, while becomes constant far from it. It is difficult to do a fair comparison between $r_{2D}$ with the value of the backscattering calculated using a 1D assumption ($r_{1D}$) because a question arises about the correct definition of $Q_{sat}$ in a 2D geometry. However if we evaluate $Q_{sat}$ as the sum of the charge at $x_0(y)$ in each slice (Eq. 5), we can calculate $r_{1D}$ through Eq. 3-4. Fig. 2 shows that the value of the backscattering ratio obtained with a 1D method ($r_{1D}$) can be significantly different from the real 2D value ($r_{2D}$).

## III. Lochtefeld Method for Backscattering Extraction

In the Lochtefeld method [10, 11] to extract the backscattering coefficient in a short-channel device, the saturation inversion charge is estimated by integration of the gate-to-channel capacitance ($C_{GC}$)

$$Q_{sat} = \int_{-\infty}^{V_G+\Delta V_G} C_{GC}(V)dV \qquad (8)$$

where $V_G$ is the gate voltage of the specified bias point and $\Delta V_G$ is a correction term. Because it is difficult to measure $C_{GC}$ in a short channel device due to parasitic capacitances (overlap and instrumentation), $C_{GC}$ is measured in a longer reference device so that $\Delta V_G$ includes corrections for the threshold voltage ($V_T$) roll-off and for the DIBL. As stated in the introduction, the DIBL is evaluated in the sub-threshold regime where it is easily calculated as a simple shift of the gate voltage for a constant drain current. Figure 3 shows the barrier lowering simulated (expected DIBL) as a function of the gate bias point. This number has been calculated by taking the difference of the potentials at the virtual source and at the maximum of the charge (~1nm far from the interface) for the cases $V_D$=1V and $V_D$=50mV where $V_D$ is the drain to source voltage. It is apparent that the DIBL is a strong function of the bias point, and in particular, it is totally different in sub-threshold with respect to the inversion regime affecting the calculated inversion charge. This is due to the bias dependence of the capacitive channel-drain coupling, which is a component of the bulk capacitance. In fact the shape of the DIBL as function of the gate voltage resembles the shape of the gate capacitance as function of the gate voltage. However some authors have reported that a DIBL in the sub-threshold regime is sometime sufficient to reproduce device characteristics [21]. Once that $Q_{sat}$ is estimated, $\eta_{sat}$ is calculated from

$$Q_{sat} = qN_{2D}\Im_0(\eta_{sat}) \qquad (9)$$

From the knowledge of $\eta_{sat}$ and measuring the saturation drain current, the backscattering is extracted by Eq. 3. Let us note that Eq. 9 contains the relationship between charge and potential when $V_D$=0, while the correct equation that should be used is Eq. 4, in fact Eq. 4 reduces to Eq. 9 when $r_{sat}$=1.

# IV. Proposed Method for Barrier Lowering and Backscattering Extraction

The proposed method is based on directly using equations (3) and (4) which define the backscattering model. In addition, the term $Q_{sat}$ is estimated as

$$Q_{sat} = Q_0 + \int_{V_G}^{V_G + \frac{kT}{q}\Delta\eta} C_{GC}(V)dV \qquad (10)$$

where $\Delta\eta = \eta_{sat} - \eta_0$, $\eta_0$ is the value of $\eta$ at equilibrium ($V_D$=0) extracted from $Q_0 = qN_{2D}F_0(\eta_0)$. $Q_0$ is the equilibrium charge extracted from CV measurement corrected for the $V_T$ roll-off. Substituting Eq. 10 into Eq. 4, one obtains an equation in the two unknowns $\Delta\eta$ and $r_{sat}$:

$$Q_0 + \int_{V_G}^{V_G + \frac{kT}{q}\Delta\eta} C_{GC}(V)dV = qN_{2D}\frac{1 + r_{sat}}{2}\Im_0(\eta_0 + \Delta\eta) \qquad (11)$$

By solving the non linear system [(3) and (11)] one obtains both $r_{sat}$ and $\Delta\eta$. The DIBL is simply represented by $kT/q \cdot \Delta\eta$. Let us note that the proposed method is fully consistent with the backscattering model because Eq. 3-4 are used. Two main differences can be found with respect to the Lochtefeld method: *i)* the DIBL correction in the charge calculation is done directly using the correct barrier lowering at the specified bias point in inversion regime and not in sub-threshold , *ii)* scattering is included in the charge calculation ( the term $(1+r_{sat})/2$ in Eq. 11 ).

## V. Validation by numerical simulation and measurements

A comparison of the accuracy of the Lochtefeld method and of the proposed method is made through 2D density gradient device simulation using Medici device simulator [19]. To this purpose, in Fig. 4a and 4b we compare, for different gate lengths, the expected values of the barrier lowering (calculated as discussed in Section III) and of the backscattering (calculated by Eq. 7) obtained directly from simulation, with the ones extracted by applying the Lochtefeld method and the proposed method on the simulated I-V and C-V characteristics. Moreover the DIBL calculated in the sub-threshold regime is plotted. The simulated devices are silicon n-MOSFETs with poly-Si gate, bulk doping of $10^{18}$ $cm^{-3}$ and oxide thickness $t_{ox}$=1.2nm. Equilibrium Schrodinger-Poisson simulations show that 80% of the charge is confined in the first sub-band of the unprimed ladder $\Delta_2^{(1)}$ (see insert in Fig. 6) thus justifying, in part, the one sub-band approximation. $N_{2D}$ and $v_{th}$ (Eq. 3-4) are calculated, for the considered sub-band, with $m_{DOS}=m_C=m_t=0.19m_0$ and $g$=2, where $m_t$ is the transversal effective mass and $m_0$ is the electron free mass. Let us note the excellent agreement between the values of the expected DIBL and the DIBL extracted with our proposed method. The excellent agreement in the barrier lowering extraction is maintained also by changing the gate voltage (Fig. 3). The inconsistency of the Lochtefeld method is apparent when one compares the DIBL used to calculate the charge (expected DIBL in the subthreshold range) and the DIBL calculated as $kT/q\cdot \Delta\eta$ (Lochtefeld method in Fig.3) which is, in any case, much smaller with respect to the expected value. A comparison between the backscattering extracted with the proposed method and the one extracted with the Lochtefeld method is difficult to do because both methods are intrinsically 1D while the expected backscattering is 2D as discussed in the sub Section IIA. We repeat that the comparison is not totally fair but we are confident that the value extracted with our method is more consistent because it has been extracted by directly using the Lundstrom backscattering equations (3-4). Figure 6 show the same type of comparison for the inversion charge. The expected saturation charge has been calculated as discussed in the sub Section IIA. The charge obtained from the Lochtefeld method gives a value closer to the expected value with respect to the charge obtained with the proposed method. This apparent advantage is due to the sum of two inconsistencies as mentioned in Section IV. First, the charge is calculated with a DIBL higher with respect to the expected DIBL, that is with a DIBL in the sub-threshold regime instead of a DIBL in the inversion regime. This error should produce a charge higher with respect to the expected charge. Second, Eq. 9 is used in the Lochtefed method instead of Eq. 4, that is the increase in the charge from equilibrium to saturation is considered as due to only an electrostatic effect and scattering is neglected (the term $(1+r_{sat})/2$ in Eq. 4). Because scattering lowers the charge ( $(1+r_{sat})/2<1$ ) the value obtained with the Lochtefeld method appears lower with respect to the expected value. In any case both methods are not able to reproduce well the expected inversion charge because 2D electrostatics is neglected in the Lundstrom model and because of the one subband approximation. Experimental measurements have been performed on short-channel nMOSFETs with electrical parameters similar to those used in device simulation. The behavior of the DIBL and of the backscattering ratio extracted from experiments in Figs. 7a and 7b is in good agreement with that of the same parameters obtained from simulations reported in Figs. 4a and 4b, so that we do not spend any further comment here. The gate lengths reported in Fig. 7 are the mask lengths. The $V_T$ roll-off used to calculate $Q_0$ is calculated by using the maximum trans-conductance method for the threshold voltage extraction [20]. Moreover, the term $\Delta V_G$ in Eq. 10, includes an additional correction term ($-R_S I_{D,sat}$) due to the series resistance $R_S$ [10] which is extracted by a common linear extrapolation technique [20].

# VI. Conclusion

Standard methods to extract the backscattering need the inversion charge at the virtual source which is difficult to estimate in saturated short channel devices. This charge is usually measured in a longer reference device by a CV and after it is corrected for $V_T$ roll-off and DIBL. The DIBL correction is usually done in the sub-threshold regime with the well known method of gate voltage shift for a constant drain current. Because backscattering is calculated in inversion, the DIBL should be calculated in inversion and not in sub-threshold as usually is done. In this work we have shown that using the DIBL in sub-threshold may lead to severe errors in barrier lowering extraction and hence in the backscattering and saturation inversion charge extraction. Moreover standard methods neglect the influence of scattering in the calculation of the saturation inversion charge. We have proposed a fully experimental method to extract the DIBL and hence the backscattering and the saturation inversion charge in short channel MOSFETs that is completely consistent with the backscattering model, as it must be. The proposed experimental extraction method has been validated and applied to results from device simulation and measurements on short-channel poly-Si/SiON gate nMOSFETs with gate length down to 70 nm. Moreover we proposed an extension of the backscattering model to the case of 2D geometries. We found that the backscattering is governed by the carrier transport in a few nanometers at the silicon/oxide interface and that the value of the real 2D backscattering ratio can be sensibly affected by a 1D method when bulk MOSFETs are investigated.

# Acknowledgements

This work has been partially supported by the EC 7FP through the Network of Excellence NANOSIL (Contract 216171). The authors would like to thank IMEC for providing the samples used in this work.

# Figure Captions

Figure 1

(Top) Sub-band energy profile in saturation along the channel direction $x$. The backscattering is defined as the ratio of the negative directed current ($I^-$) to the positive directed current ($I^+$) evaluated at the virtual source ($x_0$) which is the $x$ position corresponding to the maximum of the energy. (Bottom) The backscattering in a 2D geometry is evaluated by dividing the channel along the vertical $y$ direction into thin slice $\Delta y$. The 1D model is used to calculate local values for $r_{sat}$, $I^-$ and $I^+$ into each slide and the total 2D backscattering is evaluated through Eq. 7.

Figure 2

$I^-$, $I^+$, $r_{1D}$ and $r_{2D}$ calculated as a function of the vertical depth $y$ (Fig. 1). The currents $I^+$, $I^-$ have a maximum close to the interface and decrease rapidly going far from it. $r_{2D}$ is the cumulative 2D backscattering coefficient evaluated through Eq. 7 as a function of the vertical depth $y$. As expected, $r_{2D}$ is significantly affected by the charge transport in a region of few nanometers close the interface. $r_{1D}$ is the backscattering value obtained using Eq. 3-4 with $Q_{sat}$ evaluated as sum of the charge in each slide (Fig.1 and Eq.5).

Figure 3

The barrier lowering simulated as a function of the gate bias with the proposed method, the Lochtefeld method and the one expected, calculated by taking the difference of the potentials at the virtual source and at the maximum of the charge (~1nm far from the interface) for the cases $V_D$=1V and $V_D$=50mV . The simulated devices are silicon n-MOSFETs with poly-Si gate, bulk doping of $10^{18}$ $cm^{-3}$, oxide thickness $t_{ox}$=1.2nm and gate length L=70nm. The expected DIBL is a strong function of the bias point. The DIBL calculated with our method matches very well the expected DIBL while the value calculated with the Lochtefeld procedure underestimates strongly the expected value showing a strong inconsistence with the sub-threshold expected DIBL used to calculate the inversion charge.

Figure 4

Simulated barrier lowering (a) and backscattering (b) in 2D short-channel poly-Si gate nMOSFETs with $t_{ox}$=1.2 nm at the bias $V_G$-$V_{T,LONG}$=1V and $V_D$=1V, where $V_G$ is fixed and $V_{T,LONG}$ is the $V_T$ of the long reference device. It is apparent that the values extracted with the proposed method match very well with the expected values directly obtained from the simulation using the procedure discussed in Section IIA, whereas the Lochtefeld method overestimates the barrier lowering. Moreover, the DIBL extracted with the Lochtefeld method is inconsistent with the sub-threshold expected DIBL used to calculate the inversion charge.

Figure 5

Simulated backscattering with the Lochtefeld method, the proposed method and the expected value ($r_{2D}$) as discussed in Section IIA as a function of the gate overdrive.

Figure 6

Simulated inversion charge extracted with the Lochtefeld method, the proposed method and the expected value as discussed in Section IIA. The Lochtefeld method gives a value closer to the expected value with respect to the charge obtained with the proposed method. This apparent advantage is due to the sum of two inconsistencies: the wrong DIBL correction and the assumption of $r_{sat}$=1 in the charge calculation (Eq. 9). In the inset, the results of Schrodinger-Poisson simulations show that 80% of the charge is confined in the first sub-band of the unprimed ladder ($\Delta_2^{(1)}$).

Figure 7

Barrier lowering (a) and backscattering (b) extracted from measurements in short-channel poly-Si gate nMOSFETs with $t_{ox}$=1.2 nm at the bias $V_G$-$V_{T,LONG}$=1V and $V_D$=1V, where $V_G$ is fixed and $V_{T,LONG}$ is the $V_T$ of the long reference device. The observed trends are in agreement with device simulations (Fig. 4).

# Figures

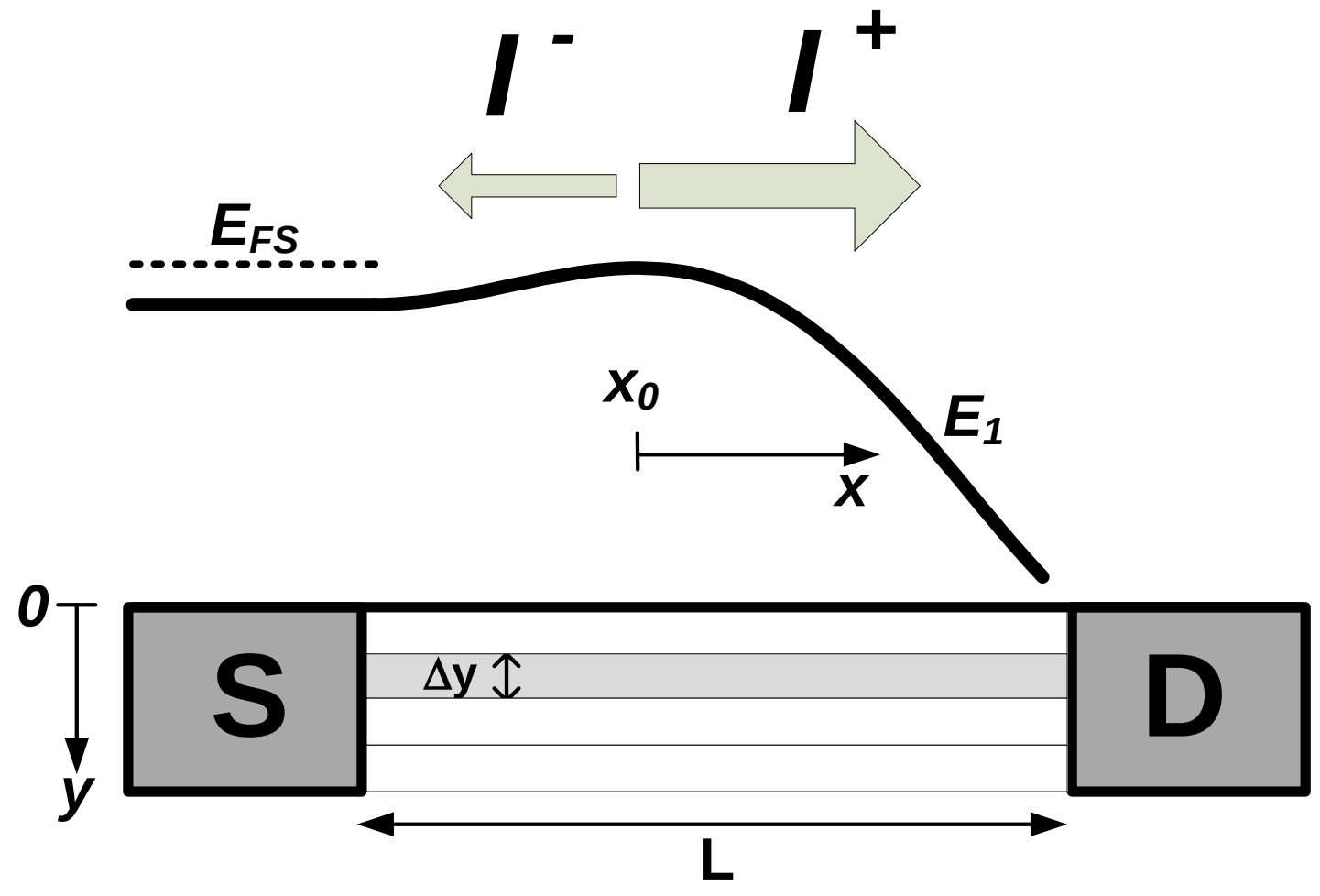

Figure 1

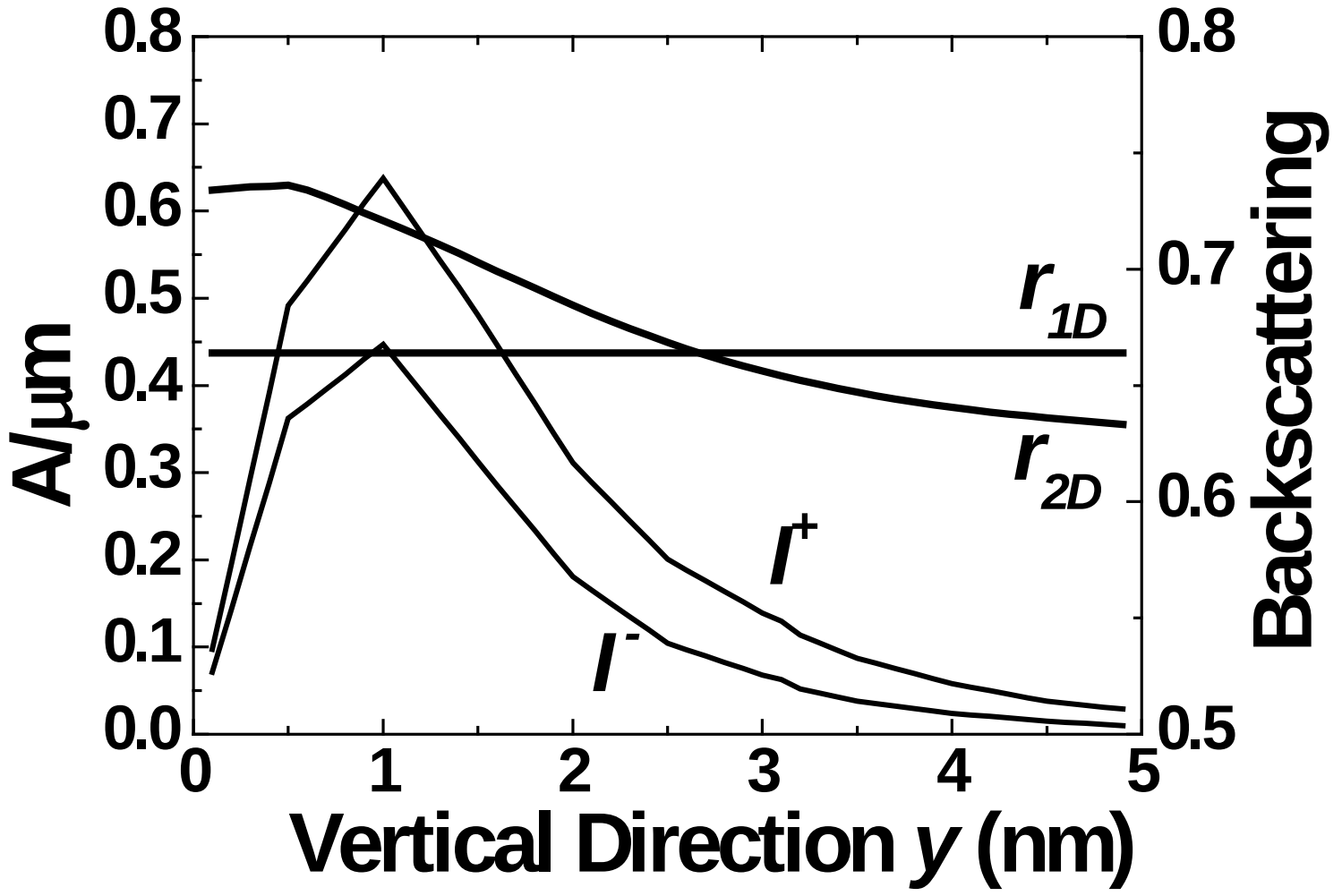

Figure 2

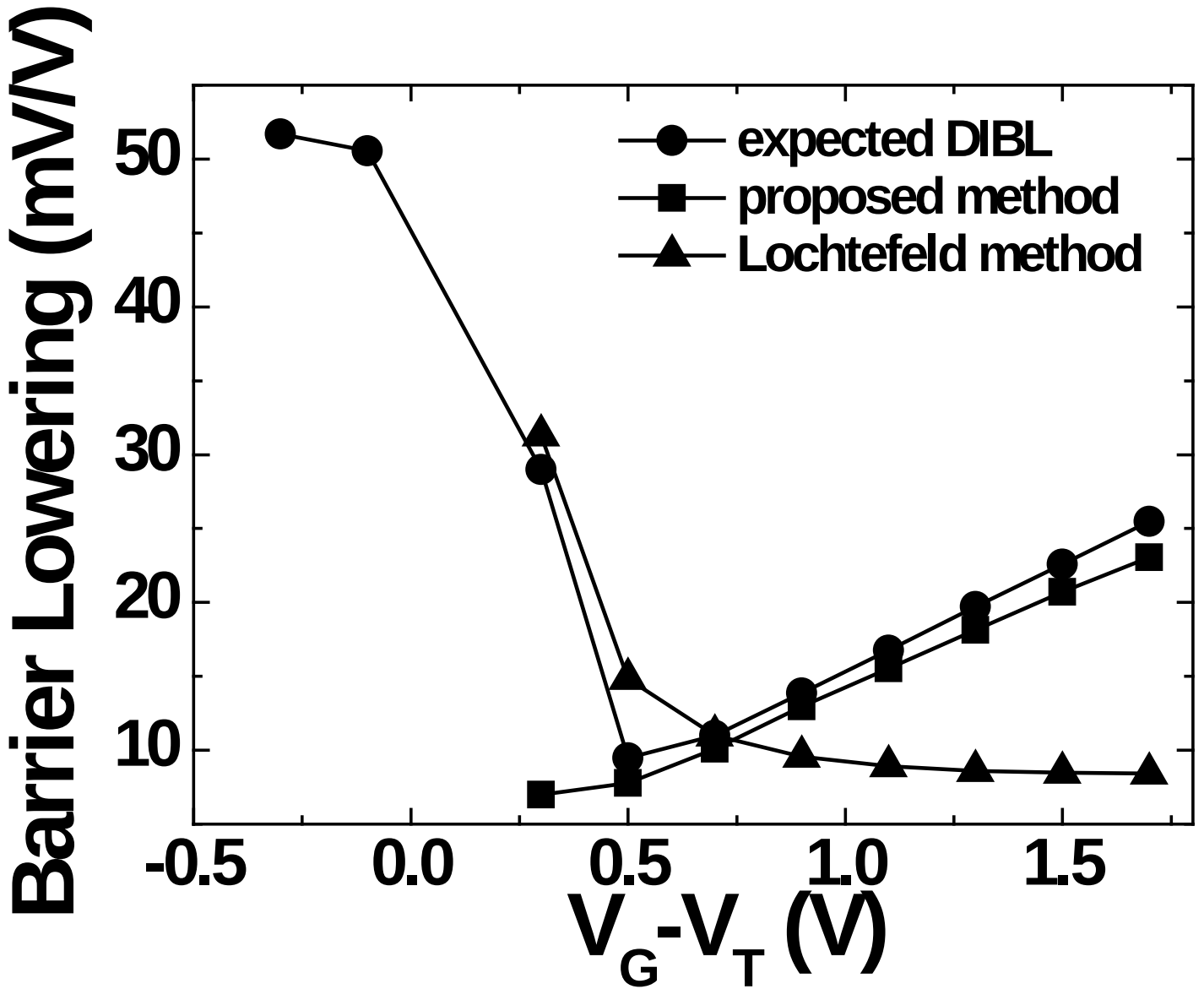

Figure 3

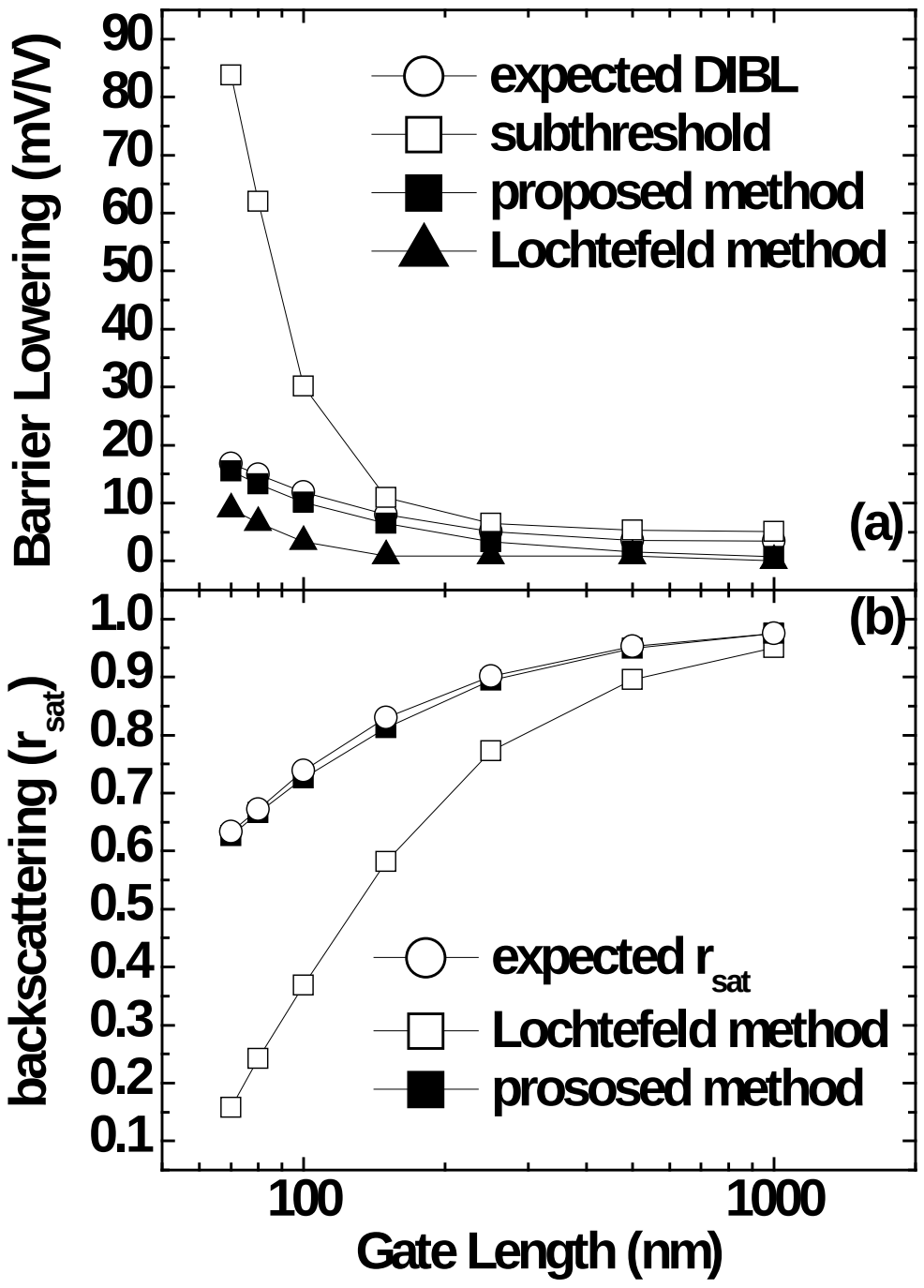

Figure 4

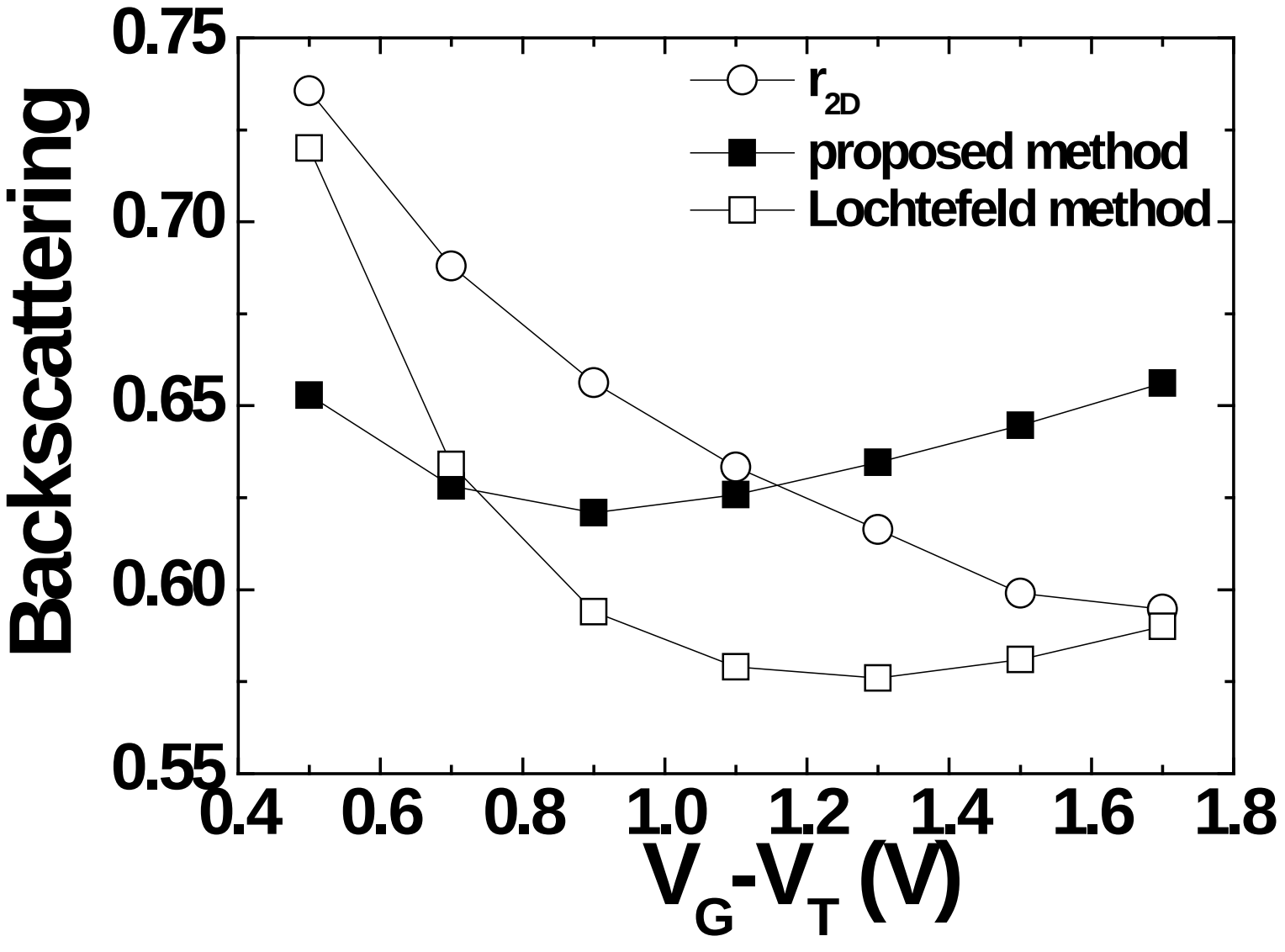

Figure 5

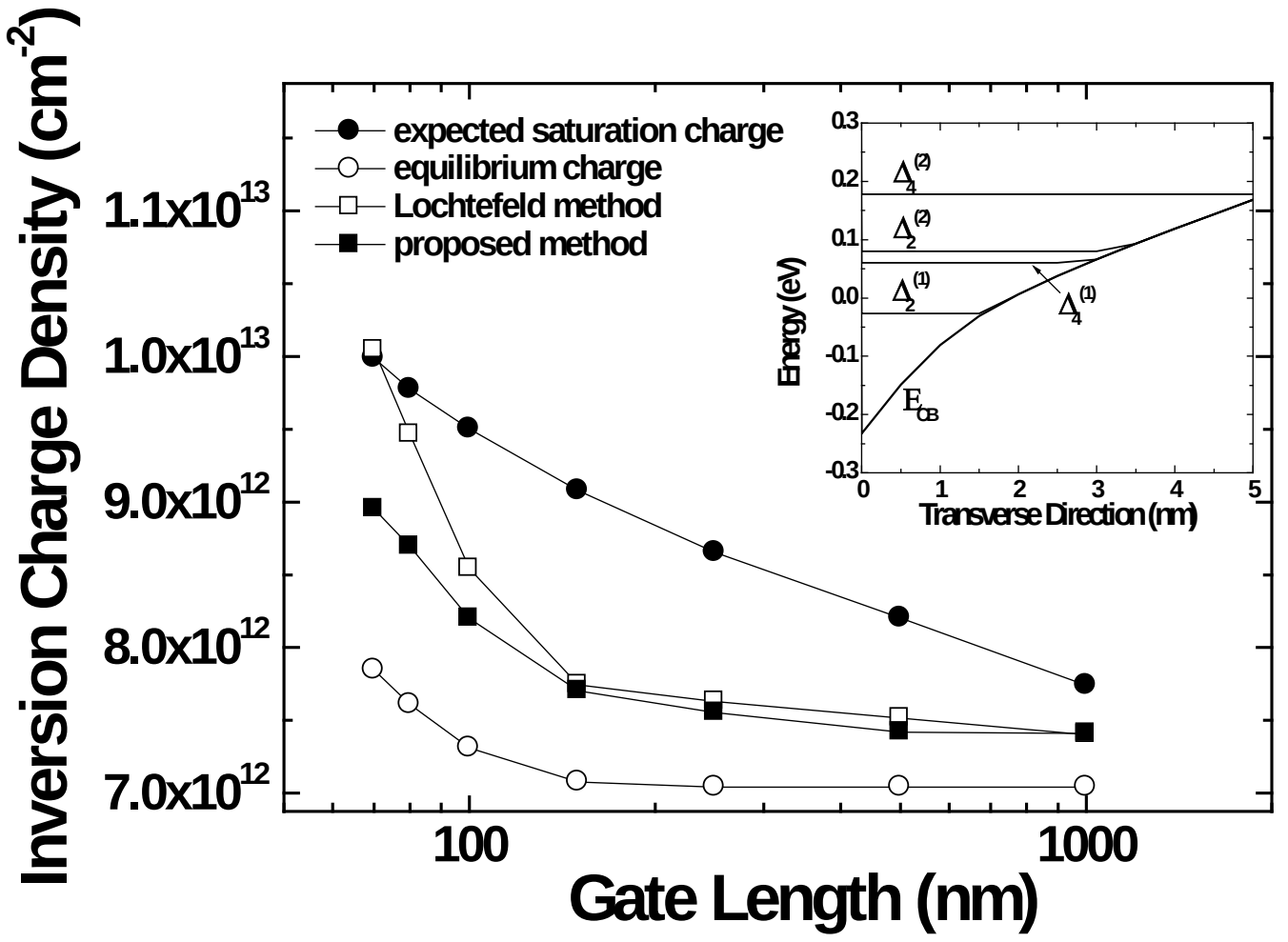

Figure 6

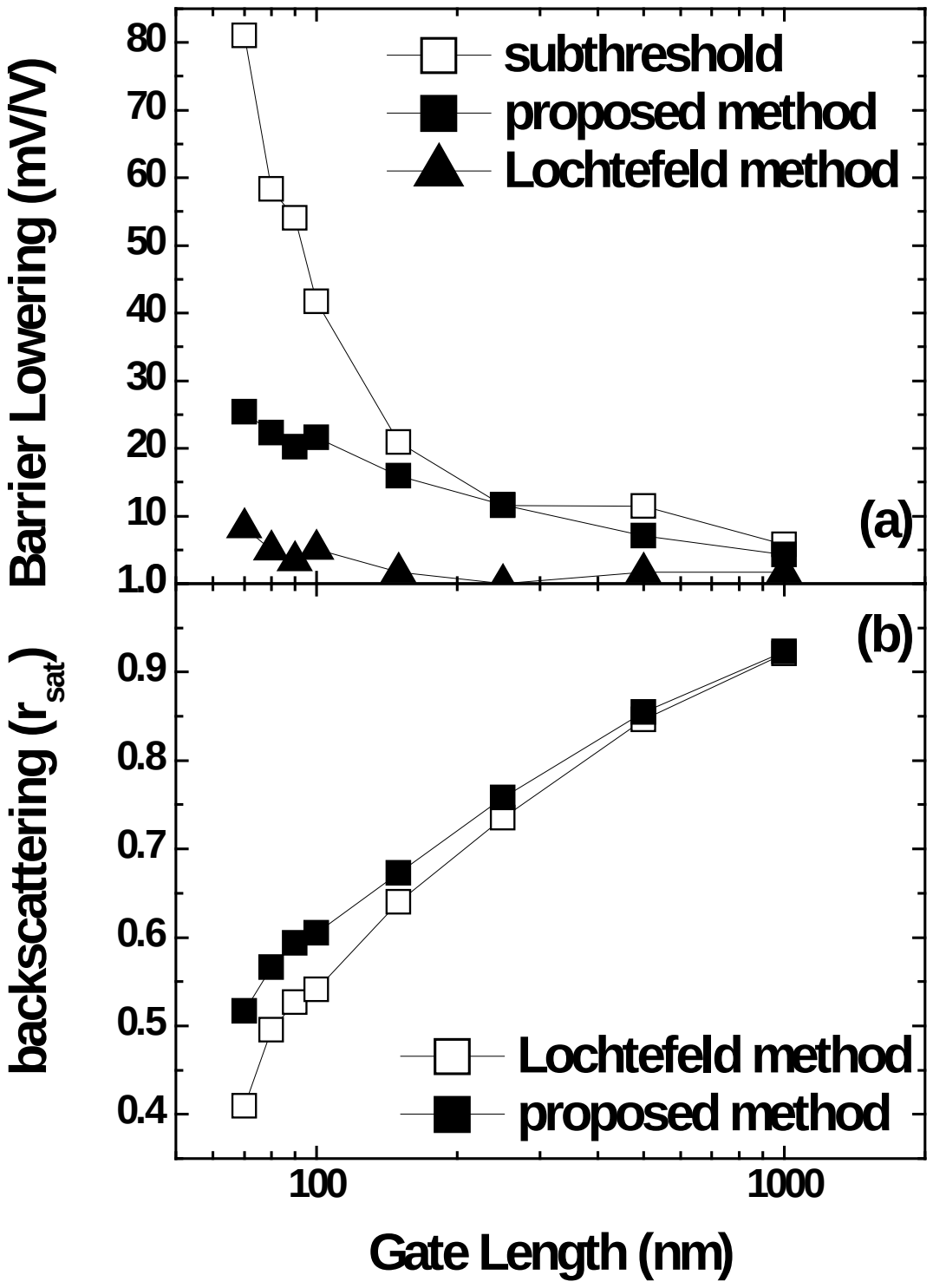

Figure 7